\pdfoutput=1

\documentclass[11pt]{article}

\usepackage[preprint]{acl}

\usepackage{times}
\usepackage{latexsym}

\usepackage[T1]{fontenc}

\usepackage[utf8]{inputenc}

\usepackage{microtype}

\usepackage{inconsolata}
\usepackage{utfsym}

\usepackage{graphicx}
\usepackage{algorithm}
\usepackage{algorithmic}

\title{Efficient Autoregressive Audio Modeling via Next-Scale Prediction}

\author{
Kai Qiu$^1$\quad Xiang Li$^1$\quad Hao Chen$^1$\quad Jie Sun$^1$\\ \textbf{Jinglu Wang$^2$\quad Zhe Lin$^3$ \quad Marios Savvides$^1$\quad Bhiksha Raj$^1$}\\
$^1$ Carnegie Mellon University\quad $^2$ Microsoft Research Asia\quad $^3$ Adobe Research\\}

\usepackage{bibentry}

\usepackage[utf8]{inputenc} 
\usepackage[T1]{fontenc}    
\usepackage{url}            
\usepackage{booktabs}       
\usepackage{amsfonts}       
\usepackage{nicefrac}       
\usepackage{microtype}      
\usepackage{xcolor}         
\usepackage{graphicx}

\usepackage{bbding}
\usepackage{graphicx}
\usepackage{amsmath}
\usepackage{amssymb}
\usepackage{booktabs}
\usepackage{colortbl}
\usepackage{bm}
\usepackage{multirow}
\usepackage{makecell}
\usepackage{tikz}
\usepackage{comment}
\usepackage{amsmath,amssymb} 
\usepackage{color}
\usepackage{colortbl}
\usepackage{tikz}
\usepackage{caption}
\usepackage{url}
\usepackage{times}
\usepackage{epsfig}
\usepackage{graphicx}
\usepackage{amsmath}
\usepackage{amssymb}
\usepackage{multirow}
\usepackage[capitalize]{cleveref}
\crefname{section}{Sec.}{Secs.}
\Crefname{section}{Section}{Sections}
\Crefname{table}{Table}{Tables}
\crefname{table}{Tab}{Tabs.}

\ExplSyntaxOn
\newcommand\latinabbrev[1]{
	\peek_meaning:NTF . {
		#1\@}%
	{ \peek_catcode:NTF a {
			#1.\@ }%
		{#1.\@}}} 
\ExplSyntaxOff

\definecolor{customgreen}{rgb}{0, 0.5, 0}
\definecolor{customred}{HTML}{E76254}
\definecolor{customblue}{HTML}{72bcd5}

\def\eg{\latinabbrev{e.g}}

\def\ie{\latinabbrev{i.e}}

\usepackage[textsize=tiny]{todonotes}

\definecolor{verylightgray}{RGB}{234, 250, 254}
\definecolor{fgreen}{RGB}{15, 159, 94}

\begin{document}

\twocolumn[{
\renewcommand\twocolumn[1][]{#1}
\twocolumn[
\maketitle]
\begin{center}
    \centering
    \vspace*{-1.2cm}
    \includegraphics[width=\textwidth]{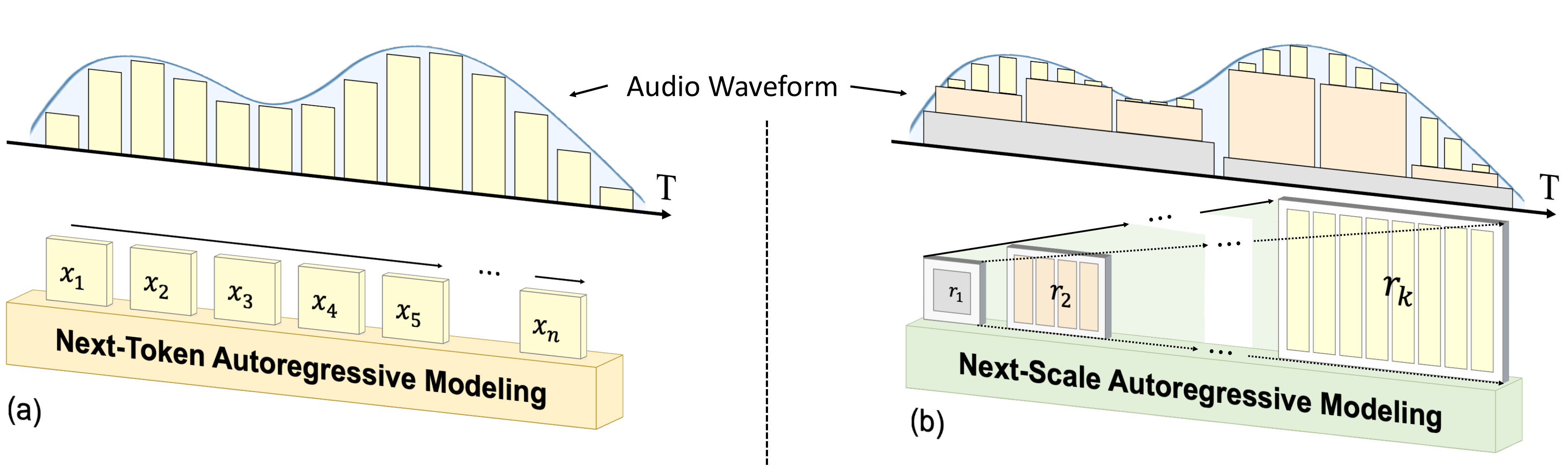}
    \vspace{-0.8cm}
    \captionof{figure}
    {Autoregressive modeling of audio. (a) Next-token prediction: sequential token generation in chronological order (left to right), which aligns with the natural temporal structure of audio; (b) Next-scale prediction: multi-scale token maps are autoregressively generated from coarse to fine scales (lower to higher resolutions). Tokens are generated in parallel within each scale, which reduces about 40x the AR prediction iteration.
    }
\label{fig:teaser}
\end{center}}
]



\begin{abstract}

Audio generation has achieved remarkable progress with the advance of sophisticated generative models, such as diffusion models (DMs) and autoregressive (AR) models. However, due to the naturally significant sequence length of audio, the efficiency of audio generation remains an essential issue to be addressed, especially for AR models that are incorporated in large language models (LLMs). In this paper, we analyze the token length of audio tokenization and propose a novel \textbf{S}cale-level \textbf{A}udio \textbf{T}okenizer (SAT), with improved residual quantization. Based on SAT, a scale-level \textbf{A}coustic \textbf{A}uto\textbf{R}egressive (AAR) modeling framework is further proposed, which shifts the next-token AR prediction to next-scale AR prediction, significantly reducing the training cost and inference time.  
To validate the effectiveness of the proposed approach, we comprehensively analyze design choices and demonstrate the proposed AAR framework achieves a remarkable \textbf{35}$\times$ faster inference speed and +\textbf{1.33} Fréchet Audio Distance (FAD) against baselines on the AudioSet benchmark. 
Code: \url{https://github.com/qiuk2/AAR}.

\end{abstract}

\def\tabnsc{
\begin{table}
\centering
\begin{tabular}{c|c|cccc}
\hline
\# scales & \# tokens & FAD & STOI & MEL & STFT\\
\hline
10 & 207 & 1.81 & 0.66 &1.55 & 2.08 \\
\hline
16 & 303 &1.52 & 0.68 & 1.46 & 1.80 \\
\hline
\hline
\end{tabular}
\caption{Performance of SAT in number of scales.}
\label{tab:nsc}
\end{table}
}

\def\tabsch{
\begin{table}
\centering
\begin{tabular}{c|c|cccc}
\hline
\hline
Scheduler & \# tokens & FAD & MEL & STFT\\
\hline
Logarithmic & 303 & 1.52 & 1.46 & 1.80\\
Quadratic & 455 & 1.40 & 1.37 & 1.86\\
Linear & 601 & 1.38 & 1.39 & 2.02\\
\hline
\hline
\end{tabular}
\caption{Ablation study on scale setting of SAT. All scale settings are trained in the same numbers of scale.}
\label{tab:sch}
\end{table}
}

\def\tabpro{
\begin{table}
\centering
\begin{tabular}{c|cccc}
\hline
Scheduler & FAD & STOI & MEL & STFT\\
\hline
Increasing & 1.38 & 0.77 & 1.39 & 2.02\\
Decreasing & 1.33 & 0.74 & 1.35 & 1.82\\
\hline
\end{tabular}
\caption{Comparison of SAT performance in enlarging and decreasing scales across different metrics.}
\label{tab:pro}
\end{table}
}

\def\tabscale{
\begin{table}
\centering
\begin{tabular}{c|c|ccc}
\hline
\hline
\# scales & \# tokens & FAD & MEL & STFT\\
\hline
10 & 207 & 1.81 & 1.55 & 2.08\\
16 & 303 & 1.52 & 1.46 & 1.80\\
\hline
\hline
\end{tabular}
\caption{Ablation study of SAT performance in number of scales.}
\label{tab:scale}
\end{table}
}

\def\tabdisc{
\begin{table}
\centering
\begin{tabular}{ccc|cccc}
\hline
\hline
STFTD & MPD & MSD & FAD & MEL & STFT\\
\hline
\usym{1F5F8} &  \usym{2717} & \usym{2717} & 1.38 & 1.36 & 1.76 \\
\usym{1F5F8} & \usym{1F5F8} & \usym{1F5F8} & 2.29 & 1.65 & 2.12 \\
\hline
\hline
\end{tabular}
\caption{Ablation study on discriminator choice. STFTD stands for Multi-scale short-time fourier transform discriminator, MPD stands for multi-period discriminator, MSD stands for multi-scale discriminator.}
\label{tab:disc}
\end{table}
}

\def\tabbs{
\begin{table}
\centering
\caption{batch size ablation.}
\begin{tabular}{c|cccc}
\hline
batch size & FAD & STOI & MEL & STFT\\
\hline
1024 & 1.38 & 0.75 & 1.36 & 1.76 \\
1536 & 1.08 & 0.75 & 1.36 & 1.75 \\
2048 & 1.20 & 0.75 & 1.35 & 1.84 \\
\hline
\end{tabular}
\label{tab:bs}
\end{table}
}

\def\tabldim{
\begin{table}
\centering
\begin{tabular}{c|cccc}
\hline
\hline
Latent dim. & FAD & MEL & STFT\\
\hline
8 & 1.47 & 1.55 & 2.15 \\
16 & 1.38 & 1.52 & 2.14 \\
32 & 1.60 & 1.43 & 2.05 \\
64 & 1.09 & 1.33 & 1.98 \\
\hline
\hline
\end{tabular}
\caption{Ablation study on latent dimension. We fix the scale to 16 and use the same quadratic scale setting. "Latent dim." represents dimension of latent representation.}
\label{tab:ldim}
\end{table}
}

\def\tabtime{
\begin{table}[h]
\centering
\begin{tabular}{c|cccc}
\hline
\hline
Window & FAD & MEL & STFT\\
\hline
1s & 1.22 & 1.36 & 1.85 \\
5s & 1.29 & 1.41 & 1.93 \\
\hline
\hline
\end{tabular}
\caption{Ablation study on temporal window.}
\label{tab:time}
\end{table}
}

\def\tabaar{
\begin{table*}[!ht]
\centering
\begin{tabular}{c|p{3cm}llll}
\hline
\hline
ID & \centering Description & FAD$\downarrow$ & IS$\uparrow$ & KL$\downarrow$ & Latency$\downarrow$ (s) \\
\hline
\rowcolor{gray!20}
1 & \centering Vanila AR & 10.05 & 2.42 & 3.01 & 7.86\\
\hline
2 & \centering AAR & 9.24$_{\textcolor{customgreen}{-0.81}}$ & 2.69$_{\textcolor{customgreen}{+0.27}}$ & 2.94$_{\textcolor{customgreen}{-0.07}}$ & 0.21$_{\textcolor{customgreen}{-7.21}}$\\
3 & \centering + Attn. Norm & 8.80$_{\textcolor{customgreen}{-1.25}}$ & 2.80$_{\textcolor{customgreen}{+0.38}}$ & 2.79$_{\textcolor{customgreen}{-0.22}}$ & 0.25$_{\textcolor{customgreen}{-7.61}}$ \\
4 & \centering + CFG & 6.44$_{\textcolor{customgreen}{-3.61}}$ & 3.52$_{\textcolor{customgreen}{+0.90}}$ & 2.32$_{\textcolor{customgreen}{-0.69}}$ & 0.25$_{\textcolor{customgreen}{-7.61}}$\\
5 & \centering + Top-k & 6.25$_{\textcolor{customgreen}{-3.81}}$ & 3.59$_{\textcolor{customgreen}{+1.17}}$ & 2.30$_{\textcolor{customgreen}{-0.71}}$ & 0.25$_{\textcolor{customgreen}{-7.61}}$\\
6 & \centering + Top-p & 6.01$_{\textcolor{customgreen}{-4.04}}$ & 3.68$_{\textcolor{customgreen}{+1.26}}$ & 2.27$_{\textcolor{customgreen}{-0.74}}$ & 0.25$_{\textcolor{customgreen}{-7.61}}$\\
\hline
\hline
\end{tabular}
\caption{Ablation study on components of AAR. vanilla AR and AAR are implemented in GPT-2 style transformer with adaptive layer normalization; "Attn. Norm" represents normalizing $q$ and $k$ into unit vector before attention; "CFG" means classifier free guidance scale of 2; Top-k and Top-p are sampling strategies where Top-k randomly selects from the top 200 indices, and Top-p (nucleus sampling) selects tokens with a cumulative probability of 0.95.}
\label{tab:aar}
\end{table*}
}

\def\tabtrain{
\begin{table}[!ht]
\centering
\resizebox{0.47\textwidth}{!}{
\begin{tabular}{c|p{1.3cm}lll}
\hline
\hline
Epoch & \centering Method & FAD$\downarrow$ & IS$\uparrow$ & KL$\downarrow$ \\
\hline
\rowcolor{gray!20}
100 & \centering AR & 7.19 & 2.78 & 2.73\\
\hline
10 & \centering AAR & 7.57$_{\textcolor{customred}{+0.38}}$ & 2.97$_{\textcolor{customgreen}{+0.19}}$ & 2.74$_{\textcolor{customred}{+0.01}}$\\
20 & \centering AAR & 6.83$_{\textcolor{customgreen}{-0.36}}$ & 3.24$_{\textcolor{customgreen}{+0.46}}$ & 2.53$_{\textcolor{customgreen}{-0.20}}$\\
30 & \centering AAR & 6.36$_{\textcolor{customgreen}{-0.83}}$ & 3.49$_{\textcolor{customgreen}{+0.71}}$ & 2.40$_{\textcolor{customgreen}{-0.33}}$\\
40 & \centering AAR & 6.32$_{\textcolor{customgreen}{-0.87}}$ & 3.55$_{\textcolor{customgreen}{+0.77}}$ & 2.32$_{\textcolor{customgreen}{-0.41}}$\\
45 & \centering AAR & 6.13$_{\textcolor{customgreen}{-1.06}}$ & 3.63$_{\textcolor{customgreen}{+0.85}}$ & 2.28$_{\textcolor{customgreen}{-0.45}}$\\
\hline
\hline
\end{tabular}
}
\caption{Comparison of training cost between vanilla AR and our AAR. All results are generated within classifier-free guidance scale of 4.}
\label{tab:train_cost}
\end{table}
}

\section{Introduction}


Autoregressive (AR) modeling \cite{achiam2023gpt, sun2024autoregressive} has been widely used in the generation domain, which typically involves two steps - 
token quantization \cite{esser2021taming, yu2021vector} and next-token prediction \cite{achiam2023gpt, touvron2023llama}. Specifically, the token quantization aims to convert the inputs to a sequence of discrete tokens and the next-token prediction models the conditional distribution of one token based on previous ones. 
AR approaches have shown significant success in textual modeling, \eg, large language models (LLMs) \cite{vaswani2017attention, devlin2018bert, touvron2023llama, achiam2023gpt} and even visual modeling \cite{dosovitskiy2020image, chang2022maskgit}. However, despite its effectiveness, AR-based audio generation remains under-explored.

Unlike natural language which is discrete and can be easily tokenized into a short series of tokens, audio demonstrates more challenges to be discretized without losing perceptual quality given its long sequence and continuity nature. Previous approaches \cite{defossez2022high, yang2023hifi, kumar2024high, zeghidour2021soundstream} leverage multi-stage residual quantization (RQ) \cite{lee2022autoregressive} to model the raw waveform with different frequencies. However, the multi-stage RQ will significantly increase the token length, leading to difficulty in the subsequent next-token prediction. Another paradigm \cite{baevski2020wav2vec} focuses on the semantics of the waveform and leverages pre-trained models (\eg, Hubert \cite{hsu2021hubert}) to cluster the embeddings in the semantic space and then quantize the embeddings based on cluster centers. Though semantic embeddings can successfully reconstruct the waveform, the reconstruction quality and generalization capability are bottlenecked by the pre-trained encoder. 

In addition, compared to text and images, audio waveform typically has a much longer sequence length due to the high sampling rates, such that about 960000 sequence length in 1 min audio clip with 16kHz. Since AR models predict tokens in a sequential manner, the inference cost is quadratically correlated to the sequence length, making the AR-based audio generation slow and computationally expensive, as illustrated in \cref{fig:teaser} (a).

VAR \cite{tian2024visual, li2024imagefolder, li2024controlvar}, a recent variant of AR models shifts from token-wise to scale-wise AR prediction scheme with a multi-scale tokenizer, showcasing improved efficiency and scalability in visual domains. 
However, applying scale-wise prediction to raw audio generation remains challenging due to the high temporal resolution of audio signals, making efficient audio generation difficult with existing methods. Large token length from the tokenizer will burden the AR modeling. Unlike the 2D visual tokenizer that compresses images in spatial (vertical and horizontal) axes, the audio tokenizer typically only compresses along the temporal axis, making it challenging to achieve a high compression rate. To address this issue, we leverage a trade-off between token length and the residual depth. Specifically, multi-scale quantization reduces the token number in each scale, allowing for more residual layers under the same total token constraint, thereby enhancing performance. This observation highlights the potential of multi-scale designs for tokenizer optimization, especially for audio applications.

In this paper, we explore the \textbf{S}cale-level \textbf{A}udio \textbf{T}okenizer and Multi-Scale \textbf{A}coustic \textbf{A}uto\textbf{R}egressive Modeling in audio generation, as illustrated in \cref{fig:teaser} (b). On the one hand, to shorten the audio token length, we utilize a scale-level audio tokenizer (SAT) which improves the traditional residual quantization with a multi-scale design and compresses the token length according to the scale index. On the other hand, we further shorten the inference step during the autoregressive prediction. Based on the multi-scale audio tokenizer, we propose acoustic autoregressive modeling (AAR) which models the audio tokens with a next-scale paradigm. Since each scale contains multiple audio tokens, the AAR can lead to much fewer autoregressive step numbers during inference compared to the traditional token-level modeling. By reducing both the token length and the autoregressive step number, our approach achieves not only a superior generated audio quality but also a remarkably faster (about $35\times$) inference speed. 

Our contribution is three-fold:
\begin{itemize}
\item We present \textbf{S}cale-level \textbf{A}udio \textbf{T}okenizer (SAT) for audio reconstruction which can efficiently compress audio sequence to tokenizers with different scales.
\item Based on SAT, we introduce scale-level \textbf{A}coustic \textbf{A}uto\textbf{R}egressive modeling (AAR), significantly reducing the inference latency and training cost.
\item Extensive experiments are conducted to analyze the performance of the proposed approach, providing insights into its capabilities and potential applications in the field of audio synthesis. 
\end{itemize}

\begin{figure*}[t]
  \centering
  \includegraphics[width=\linewidth]{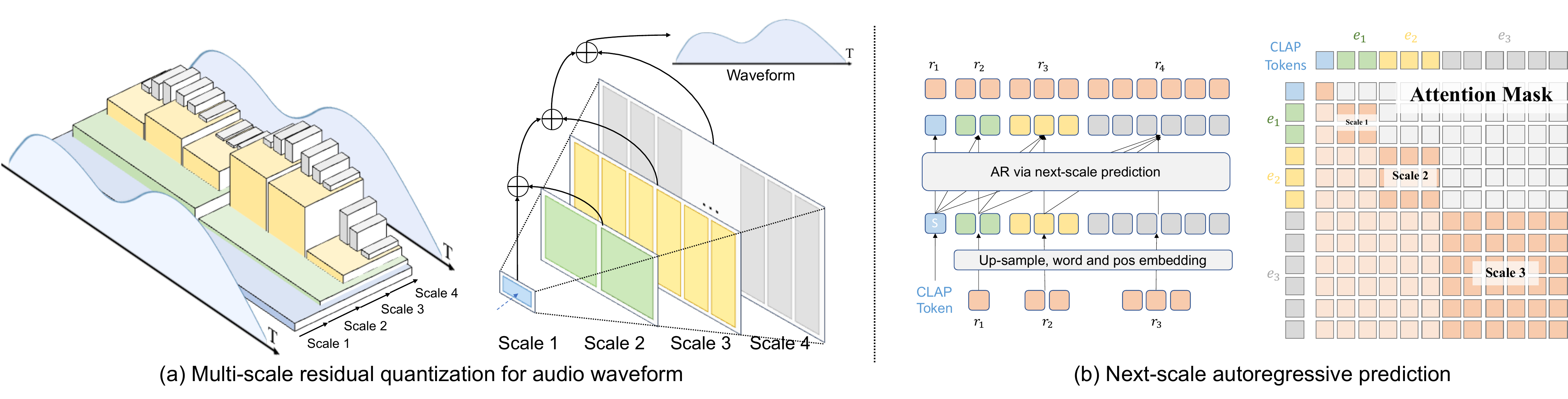}
  \caption{
  Our model involves two distinct training phases. \textbf{Stage 1}: Scale-level Audio Tokenizer (SAT) to encode an audio sample into a series of $K$ tokens scales, donated as $\mathcal{R} = (r_1, r_2, \dots, r_K)$. Each scale encodes information in different frequencies of the audio waveform.
  \textbf{Stage 2}: Acoustic AutoRegressive (AAR) modeling via next-scale prediction relies on the pre-trained SAT to predict each scale-level token $r_i$ by conditioning on all previously predicted scales $r_{<i}$ and a CLAP token \cite{wu2023large} as the start token. The CLAP token is derived from ground truth audio. During training, we use the standard cross-entropy loss and the attention mask as figured above to ensure that each $r_i$ can only be attributed by $r_{\leq i}$ and the start token.
}
  \vspace{-0.4cm}
  \label{fig:pipe}
\end{figure*}

\section{Related Works}

\paragraph{Raw audio discretization.}
Before the development of Variational Autoencoders (VAEs) \cite{van2017neural,razavi2019generating}, converting continuous domains into discrete representations was a significant challenge in generative modeling. VAEs facilitate the effective quantization of inputs into structured priors using powerful encoder-decoder networks, allowing manipulation in tasks like generation and understanding \cite{achiam2023gpt,touvron2023llama,caillon2021rave}. Recent innovations, such as VQGAN \cite{esser2021taming} and RQGAN \cite{lee2022autoregressive}, have further advanced these priors, improving model generalization and inspiring numerous works in audio discretization \cite{oord2016wavenet, caillon2021rave,siuzdak2024snac, li2024xq}. In the audio domain, Encodec \cite{defossez2022high} employs an architecture similar to SoundStream \cite{zeghidour2021soundstream}, using an encoder-decoder model to reconstruct audio, incorporating residual quantization and a spectrogram-style discriminator to enhance audio quality. In contrast, HIFI-codec \cite{yang2023hifi} uses group residual quantization to refine the representation in the initial quantization layer. Kumar et al. \cite{kumar2024high} have made significant contributions to audio reconstruction by introducing multi-spectrogram loss and quantizer dropout to enhance bitrate efficiency. Building upon these advances, our work poses an important question: can we use fewer tokens to represent low-frequency information, thereby efficiently reducing the burden while maintaining high-quality audio reconstruction? To address this, we propose a \textbf{S}cale-level \textbf{A}udio \textbf{T}okenizer, which encodes audio on different scales, capturing hierarchical features that improve both the efficiency and quality of audio generation and reconstruction.


\paragraph{Autoregressive modeling}
The autoregressive model \cite{chowdhery2023palm,hoffmann2022training}, as a different approach from diffusion models, leverages efficient Large Language Models (LLMs) \cite{vaswani2017attention, devlin2018bert, touvron2023llama, achiam2023gpt, kreuk2022audiogen, wu2023ar} to generate the next tokens sequentially to construct the output. Due to its sequential nature, autoregressive models have excelled in text generation, machine translation, and other sequence prediction tasks. Recently, autoregressive models have also made significant processes in the image generation domain \cite{chang2022maskgit, sun2024autoregressive}. By treating image pixels or patches as sequences, these models can generate high-quality images by sequentially predicting each part of the image.


\section{Preliminary: Vanilla Audio Tokenizer}

\paragraph{Audio quantization.}

Consider an audio signal $a\in \mathbb{R}^{C\times T}$, where $C$ represents the number of audio channels and $T$ is the number of samples over the duration of the signal. Traditional approach \cite{kumar2024high,defossez2022high,yang2023hifi} in audio tokenizer often involves a 1D convolutional-based autoencoder frameworks to compress audio waveform to latent space $x\in \mathbb{R}^{l\times d}$ where $l$ is the token length and then utilizes a vector quantization to quantize the latent tokens:
\begin{equation}
    x = \mathcal{E}(a),\ \ \ \ \hat{x} = \mathcal{Q}(x),\ \ \ \ \hat{a} = \mathcal{D}(\hat{x})
\end{equation}
where $\mathcal{E}(\cdot)$ donates encoder, $\mathcal{Q}(\cdot)$ a vector quantizatier, and $\mathcal{D}(\cdot)$ a decoder. A vector quantizer $\mathcal{Q}$ maps each feature vector in the latent space $x$ to the closest vector in a learnable codebook $Z \in \mathbb{R}^{d\times V}$ with $V$ vectors of dimension $d$. Specifically, vector quantization $\hat{x} = \mathcal{Q}(x)$ involves looking up the closest match for each feature vector in $x$ with vectors in $Z$ by minimizing Euclidean distance: 
\begin{equation}
    \hat{x} = \text{argmin}_{z\in Z} || x - z ||_2
\end{equation}
where $\hat{x}$ represents the quantized output and $x$ is the input to the quantizer. 

However, due to the complexity of the audio waveform, particularly in handling frequency-specific information, a residual quantization approach is typically employed. In residual quantization, a sequence of vector quantizers $\mathcal{Q}=\{\mathcal{Q}_1,\mathcal{Q}_2,\cdots,\mathcal{Q}_K\}$ is used, where each quantizer $\mathcal{Q}_i$ iteratively quantizes the residual error from the previous step. Specifically, after each quantization step, the residual error is computed as $\delta_i = x_i - \hat{x_i}$ and passed to the next quantizer as the input $x_{i}=\delta_{i-1}$. The final quantized representation $\hat{f}$ is obtained by summing the outputs from all quantizers
\begin{equation}
    \hat{x}=\sum_{i=1}^{r}\hat{x_i}
\end{equation}
which is then decoded by the decoder $\mathcal{D}(\hat{x})$ to produce the reconstructed output $\hat{a}$.

\paragraph{Loss function.}
To train audio quantized autoencoder, we leverage a combination of loss functions including the reconstructed time-domain loss $\mathcal{L}_t$, reconstructed frequency domain loss $\mathcal{L}_f$, discriminative loss $\mathcal{L}_G$, residual quantization loss $\mathcal{L}_{vq}$, and commitment loss $\mathcal{L}_{commit}$ \cite{defossez2022high} :

\begin{equation}
    \mathcal{L} = \lambda_t\mathcal{L}_t + \lambda_f\mathcal{L}_f + \lambda_G\mathcal{L}_G + \mathcal{L}_{vq} + \lambda_{com}\mathcal{L}_{com}.
\end{equation}
Specifically, reconstructed time-domain loss measures the absolute difference between $a$ and $\hat{a}$ as
\begin{equation}
    \mathcal{L}_t = ||a-\hat{a}||
\end{equation}
and frequency domain loss assesses the difference over mel-spectrograms across $n$ time scales as 
\begin{equation}
    \mathcal{L}_f = \sum_{i=1}^n||\mathcal{S}_i(a)-\mathcal{S}_i(\hat{a})|| + ||\mathcal{S}_i(a)-\mathcal{S}_i(\hat{a})||_2^2
\end{equation}
where $\mathcal{S}_i$ represents the transformation to the mel-spectrogram at scale $i$. The discriminative loss is derived from a multi-scale STFT discriminator, as introduced in \cite{zeghidour2021soundstream} to ensure the model captures high-fidelity audio features across various time-frequency scales.
The vector quantization loss encourages the encoded features to match the codebook vectors, and the commitment loss penalizes deviations from these vectors, ensuring that the encoder commits to the quantized space as
\begin{equation}
\begin{aligned}
    \mathcal{L}_{vq} = \sum_{i=1}^r||\text{sg}(x_i)-z_i||_2^2\\
    \mathcal{L}_{com} = \sum_{i=1}^r||x_i - \text{sg}(z_i)||_2^2.
\end{aligned}
\end{equation}

\paragraph{Analysis.}
The baseline audio tokenizer can successfully discretize audio tokens. However, due to the residual quantization, the token length representing each audio will be significant which severely hinders the efficiency in the autoregressive modeling. Considering each quantizer in residual quantization basically divides and represents the audio into different frequency bands \cite{defossez2022high,kumar2024high,yang2023hifi}, we aim to further adjust the token length based on its represented frequencies, \ie, lower-frequency parts can be represented with fewer tokens. To this end, we introduce the \textbf{S}cale-level \textbf{A}udio \textbf{T}okenizer to reduce the number of tokens being used.

\section{Method}


Our approach consists of two major stages: (1) In the first stage, we train a \textbf{S}cale-level \textbf{A}udio \textbf{T}okenizer (SAT) to convert continuous audio signals into discrete tokens using multi-scale residual quantization. (2) The second stage reformulates the \textbf{A}coustic \textbf{A}uto\textbf{R}egressive modeling (AAR) in a next-scale manner and models the tokens obtained from the frozen SAT tokenizer with a transformer structure.


\subsection{Scale-level Audio Tokenizer} 
In Scale-level Audio Tokenizer (SAT), we employ the same encoder-decoder architecture as baseline tokenizer \cite{defossez2022high} but incorporate multi-scale residual quantization (MSRQ) to enhance efficiency and flexibility in audio representation. In MSRQ, as shown in \cref{fig:pipe} (a), the quantizer $\mathcal{Q}_i$ is defined the same as the baseline setting as $r_i=\mathcal{Q}_i(r_{i-1})$ while the feature map $r_{i-1}$ is first downsampled from its original dimension $l_K\times d$ to a lower resolution $l_k\times d$ where $K$ is the scale number of the last index and $k$ is the scale number of the correct index. After downsampling, the look-up procedure is performed to match each feature vector with the closest codebook vector $Z_i$. After the look-up, the processed quantized vector $z_i$ is upsampled back to the original dimension $l_K\times d$ to ensure consistency across scales. Due to the loss of high-frequency information from downsampling, we employ a 1D convolutional layer after upsampling to restore the lost details and enhance the fidelity of the reconstructed audio. Specifically, this convolutional layer processes the upsampled feature vectors according to the equation
\begin{equation}
    \phi(\hat{r}) = \gamma\times\text{conv}(\hat{r}) + (1-\gamma)\times \hat{r}
\end{equation}
where $\text{conv}(\cdot)$ 
applies a 1D convolution with a kernel size of 9. This design effectively combines the original features with the transformed outputs, while preserving the reparameterization inherent to vector quantization, controlled by the quantization residual ratio $\gamma$. 
In the Appendix, we provide a pseudo-code for the scale-level audio tokenizer.

\subsection{Acoustic AutoRegressive Modeling}


\paragraph{Vanilla autoregressive modeling.}
Autoregressive modeling is first introduced by \cite{sutskever2014sequence, bahdanau2014neural}  and quickly spread to different modalities such as image \cite{sun2024autoregressive}, video \cite{weissenborn2019scaling} and 3D modeling \cite{siddiqui2024meshgpt}. In autoregressive modeling, a sequence of data points is modeled as a product of conditional probabilities. For a sequence $x = (x_1, x_2, \dots, x_T)$, its joint distribution can be expressed and modeled as  
\begin{equation}
p(x_1, x_2, ..., x_T) = \prod_{t=1}^{T} p(x_t | x_1, x_2, ....x_{t-1}).
\end{equation}
This approach is widely used across various domains due to its flexibility and ability to capture dependencies within data. For any continuous modality, it is traditional to first train a tokenizer to discretize the input into tokens, which can then be modeled using a discrete categorical distribution. This step involves mapping the continuous data to a sequence of discrete tokens $x = (x_1, x_2, \dots, x_T)$ that are fed into an autoregressive model to predict the next token in the sequence, based on the preceding tokens.
In the context of transformers, which have become the dominant architecture for autoregressive modeling, the attention mechanism plays a crucial role in training. The attention mechanism allows the model to focus on different parts of the input sequence when making predictions. To ensure that the model adheres to the autoregressive property, where each token $x_i$ is predicted based only on previous tokens $x_1, x_2, \dots, x_{i-1}$, an attention mask is applied. Mathematical, the attention mask $M$ is defined as 
\begin{equation}
    M_{ij} = 
    \begin{cases} 
    1, & \text{if } i \leq j \\
    0, & \text{otherwise}
    \end{cases}
\end{equation}
where guarantee the modeling's performance on predicting $x_i$ is only relevant to its preceding tokens.

After the completion of training of such a model $P$ using cross-entropy loss, it can efficiently handle complex dependencies and generate new samples by sequentially predicting each token conditioned on its predecessors \cite{achiam2023gpt,touvron2023llama,sun2024autoregressive}.

This capability makes autoregressive models well-suited for generating data that requires a coherent and consistent sequence. However, their capacity for audio generation is still under-explored due to the huge sequence length required for audio data. The sheer number of tokens needed to represent even short audio clips can lead to computational inefficiencies and challenges in maintaining temporal coherence. To efficiently solve such a challenge, we combine the unique property of our SAT to efficiently generate audio via scale-level \textbf{A}coustic \textbf{A}uto\textbf{R}egressive modeling.

\begin{table*}
\centering
\scalebox{1}{
\begin{tabular}[t]{
c|l|c|p{0.9cm}<{\centering}p{0.9cm}<{\centering}p{0.9cm}<{\centering}|p{0.9cm}<{\centering}p{0.9cm}<{\centering}} 
\hline
\hline
\multirow{2}*{ID} & \multirow{2}*{Method} & \multirow{2}*{Dataset} & \multicolumn{3}{c|}{Reconstruction} & \multicolumn{2}{c}{Generation}\\
\cline{4-8}
 ~ & ~ & ~ & $|\mathcal{L}|$ & rFAD$\downarrow$ & MEL$\downarrow$ & gFAD$\downarrow$ & KL$\downarrow$ \\
\hline
\rowcolor{gray!20}
1 & Vanilla AR (baseline) & AS & 750 & 1.39 & 1.33 & 10.05 & 3.01 \\
\hline
2 & DiffSound \cite{yang2023diffsound} & AS & - & 
- & - & 9.76 & 4.21 \\
3 & AudioLCM \cite{liu2024audiolcm} & AC + LP & - & - & - & 3.92 & 1.20 \\
4 & AudioLDM 2 \cite{liu2024audioldm} & AS + 7 more & - & - & - & 1.42 & 0.98\\
\hline
5 & AAR (ours) & AS & 455 & 1.09 & 1.33 & 6.01 & 2.27 \\
\hline
\hline
\end{tabular}
}
\caption{We evaluated the performance of our AAR model against other methods using rFAD and MEL Distance to measure reconstruction quality, and gFAD and KL Divergence to assess generation quality, where $\downarrow$ indicates that lower values are better. In this context, $|\mathcal{L}|$ denotes the token length. Additionally, AS, AC, and LP denote the datasets AudioSet, AudioCaps, and LP-Musicaps, respectively.
 }
\label{tab:ablation}
\end{table*}

\paragraph{Acoustic autoregressive modeling.}
To shorten the inference step, we propose Acoustic autoregressive modeling (AAR). This approach, distinct from traditional vanilla autoregressive models that predict token sequences one by one, involves predicting across different scales. Attributed by SAT, our method represents an audio sample as a series of scale-level representations:
\begin{equation}
    \mathcal{R} = (r_1, r_2, \dots, r_K)
\end{equation}
By efficiently expressing it as joint modeling, the audio sequence is defined as:
\begin{equation}
    p(\mathcal{R}) = \prod_{i=1}^{K} p(r_i | r_1, r_2, ...,r_{i-1})
\end{equation}
In this formulation, each $r_i$ represents a distinct scale in the hierarchical representation of the audio signal. The model predicts each scale by conditioning on all previously predicted scales, effectively capturing both global structures and fine-grained details of the audio. This hierarchical approach reduces the complexity associated with long sequence lengths by leveraging multi-scale dependencies, thereby enhancing the model's efficiency and ability to maintain temporal coherence. To successfully implement our method, we modify the attention mask $M$ for each scale $r_i$ to focus only on the relevant scales:
\begin{equation}
    M_{ij} = 
    \begin{cases} 
    1, & \text{if } i \leq j \\
    0, & \text{otherwise}
    \end{cases}
\end{equation}
This attention mask ensures that the model only attends to $r_1, r_2, \dots r_{i-1}$ when predicting $r_i$ ignoring future scales and reducing unnecessary computations. A detailed description of the implementation is summarized in the Appendix.


\section{Experiment}

\begin{figure}[t]
  \centering 
  \includegraphics[width=\linewidth]{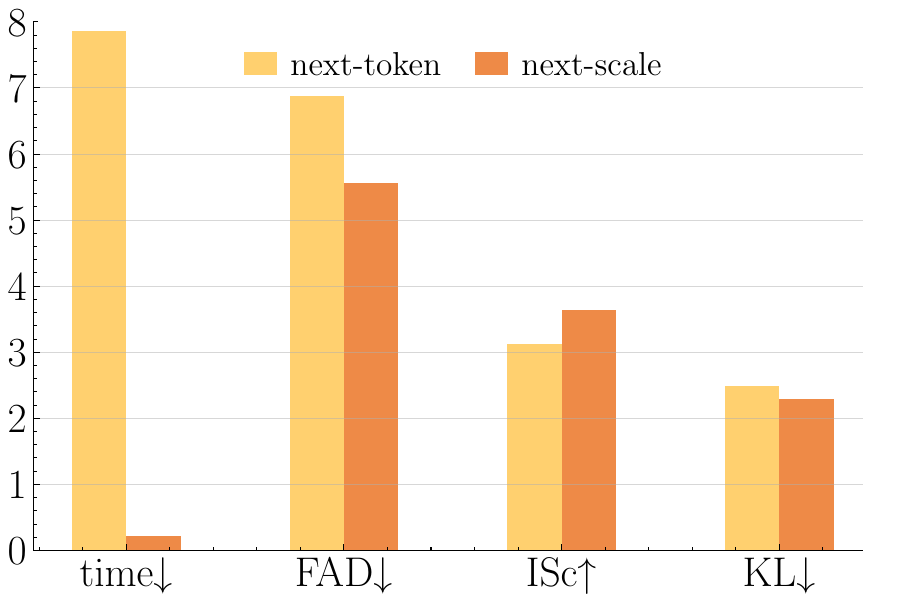}
  \caption{Performance of autoregressive model when classifier-free guidance is 10. next-token: AR via next-token prediction; next-scale: our AAR.}
  \label{fig:exp2}
  \vspace{-0.5cm}
\end{figure}

\subsection{Evaluation Metrics and Settings}
We evaluate FAD \cite{kilgour2018fr}, MEL distance \cite{kumar2024high}, and STFT distance \cite{kumar2024high} as reference for reconstruction, and FAD \cite{kilgour2018fr}, ISc and KL \cite{salimans2016improved} for generation. FAD, built upon VGGish \cite{chen2020vggsound}, is the metric to indicate the similarity of the generated and target samples effectively. MEL distance quantifies the difference in mel-spectrogram features, and STFT distance measures the short-time Fourier transform discrepancies between the generated and target audio signals, which focus more on high-frequency information for audio. Additionally, ISc, simulating its performance on image generation, is used to evaluate the generated sample diversity and quality. KL divergence is utilized to measure the difference between the probability distributions of the generated and target samples.

We conducted all experiments on the AudioSet \cite{gemmeke2017audio} dataset. To effectively evaluate the performance of our audio tokenizer, we divided the original 10-second evaluation set into $n$ segments, each matching the window size of our model for reconstruction. After reconstructing these segments, we reassembled them into a complete audio stream. For autoregressive generation, we randomly selected one segment from the evaluation set and used it as the ground truth. 


\subsection{Implementation Details}

\paragraph{Tokenizer.}
In stage 1, we utilize multi-scale residual quantization  (MSRQ) of codebook size 1024 with the Soundstream autoencoder framework \cite{zeghidour2021soundstream}.  The model is trained for 100 epochs using the Adam optimizer with $\beta_1 = 0.5$ and $\beta_2=0.9$. We apply a cosine learning rate scheduler with initial learning rate 3e-4 and set the loss weights to $\lambda_t=0.1, \lambda_f=3,\lambda_G=3, $ and $\lambda_{com}=1$. Our discriminator updated 2/3 times during training.

\paragraph{Transformer.}
In stage 2, our primary focus is on scale-level acoustic autoregressive modeling. To achieve this, we employ a GPT-2-style transformer \cite{radford2019language} with adaptive normalization \cite{zhang2018self} and depth of 16. We utilize CLAP audio embeddings \cite{wu2023large} as our start tokens. Since one-second audio segments often contain limited meaningful information, we opt to use 10-second audio embeddings to capture richer context, even when generating one-second clips. For training, we adopt the AdamW optimizer with a learning rate of 1e-4, using a linear learning rate scheduler. Additionally, we apply a weight decay of 0.05 and implement warmup settings with an initial warmup proportion of 0.005 and an end warmup proportion of 0.01.

\subsection{Main Results Analysis}

We compare the performance of our approach with previous approaches. As shown in \cref{tab:ablation}, our proposed SAT tokenizer suppresses the baseline encodec \cite{defossez2022high} by 0.3 FAD in the reconstruction task, despite using fewer tokens (750 tokens v.s. 455 tokens). This shows that by increasing quantization while reducing the number of tokens, we can efficiently improve reconstruction quality while using fewer tokens. 


For the audio generation, we introduce an autoregressive model with next-token prediction as the baseline. To ensure a fair comparison, we employ two encoders (Encodec \cite{defossez2022high} for AR and our SAT for AAR) with similar performance. We find that our proposed AAR shows superior performance in terms of both latency and audio quality. As shown in \cref{fig:exp2}, the next-scale prediction demonstrates a remarkable improvement in audio generation, achieving a 35x speed improvement (0.225s v.s. 7.866s) and generation enhancement (FAD 5.55 v.s. 6.88).
More analysis of training costs is available in the Appendix.


\subsection{Ablation Experiments}


We conduct ablation experiments to validate the effectiveness of the components in SAT and AAR. 

\tabscale{}
\tabsch{}
\tabldim{}
\tabtime{}
\tabdisc{}

\tabaar{}

\paragraph{Effect of discriminator}
We explored multiple discriminator configurations to optimize the performance of our Scale-level Audio Tokenizer (SAT). As illustrated in \cref{tab:disc}, we tested two different discriminator setups: one using only a multi-scale short-time Fourier transform (STFT) discriminator \cite{zeghidour2021soundstream} and another combining the multi-scale STFT discriminator with a multi-period discriminator (MPD) \cite{kong2020hifi} and a multi-scale discriminator (MSD) \cite{kumar2019melgan}. Our results indicate that using only a multi-scale STFT discriminator is sufficient for effective reconstruction.

\paragraph{Effect of the scale setting.}
To find the optimal combination of SAT configuration, we start with Encodec in 128 latent dimensions with 10 quantizers \cite{defossez2022high} and test multiple scales with shared codebooks of different sizes and individual codebooks for each scale. In particular, \cref{tab:scale} shows that enlarging the scale to 16 consistently improved audio quality.
As illustrated in \cref{tab:sch}, we tested the performance of linear, quadratic and logarithmic scheduling on 16 scales: linear scheduling provides a balanced number of tokens for each scale; quadratic scheduling focuses more on the early or late stages of the process; and logarithmic scheduling offers a more gradual progression. We believe the suboptimal performance observed in logarithmic scheduling is due to its lack of high-frequency information representation at larger scales even though it also builds a complete information flow for audio. Quadratic scheduling, in particular, proved to be more efficient, requiring fewer tokens than linear scheduling (455 v.s. 601) and also achieves comparable reconstruction performance in audio quality.  

To further improve the model's capacity, we fixed the decoder dimension to 1024 and tested latent dimensions of 8, 16, 32, and 64. As \cref{tab:ldim} indicated, our SAT achieves its superior performance in the latent dimension of 64.


\paragraph{Effect of temporal windows change.} 
To effectively validate the performance of our scale scheduling in audio reconstruction, we train our SAT using 5-second audio windows with the $5\times$ original quantizer setting. This approach allows us to assess our SAT's ability to handle varying temporal dimensions and capture essential audio features over different time scales. By experimenting with different window sizes, we aim to determine the optimal configuration for maintaining high reconstruction quality while maximizing efficiency. The results of these experiments are presented in \cref{tab:time}. we find that the reconstruction quality between 1-second and 5-second windows is similar, suggesting that our SAT performs well across diverse time windows, maintaining consistent quality and demonstrating robustness in handling varying temporal scales.

\begin{figure}[]
  \centering
  \vspace{-0.2cm}
  \includegraphics[width=\linewidth]{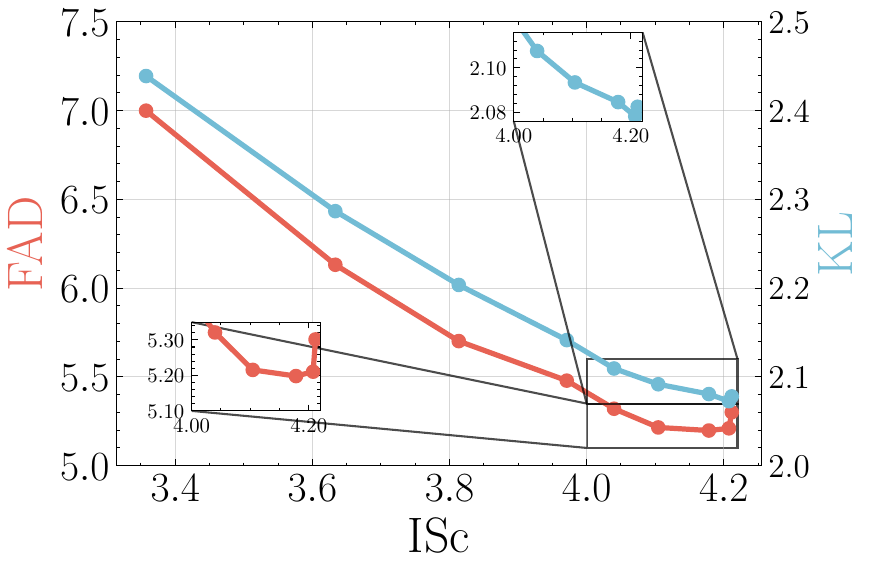}
  \vspace{-0.7cm}
    \caption{Performance of AAR in different classifier-free guidance scales from 2 to 18 (left to right), with each point incremented by 2. The \textcolor{customred}{red line} represents Fréchet Audio Distance (FAD) v.s. Inception Score (ISc), while the \textcolor{customblue}{blue line} represents Kullback-Leibler divergence (KL) vs. Inception Score (ISc).}

  \label{fig:exp1}
  \vspace{-0.3cm}
\end{figure}

\paragraph{Effect of AAR and sampling technique.}

We evaluate our AAR with the same setting as the baseline vanilla AR with roadmap shown in \cref{tab:aar}. We notice that our AAR can not only improve the generation abilities but also significantly reduce the inference time to an acceptable range. Moreover, the introduction of attention normalization can stabilize the training and further enhance the model's performance, leading to improved FAD and IS scores. The addition of CFG and advanced sampling techniques such as top-k and top-p sampling continues to push the boundaries of audio generation quality.

\paragraph{Effect of classifier-free guidance.}
As shown in \cref{fig:exp1}, we evaluate the relationship between the Inception Score with Fréchet Audio Distance and Inception Score with KL divergence across different Classifier-Free Guidance scales. We find that as the CFG scales increase, the ISc improves, while both FAD and KL metrics converge and stabilize at CFG = 14 and finally achieving FAD 5.19.


\section{Conclusion}
In this paper, we introduced a novel approach for audio generation using a multi-scale autoregressive model via next-scale prediction. This framework leverages the scale-level audio tokenizer, which efficiently compresses audio sequences into tokenizers of varying scales, thereby improving efficiency while maintaining high fidelity. Through comprehensive experiments, we demonstrated the superior performance of our method in generating high-quality audio compared to traditional autoregressive methods.

Our approach provides an efficient solution for audio generation. By incorporating a multi-scale residual quantization technique, the model effectively reduces the sequence length required for generation, leading to enhanced efficiency and reduced computational demands.

\section{Limitations.}
Despite the strong performance of next-scale prediction in general audio generation with text control, several limitations remain that warrant further exploration. Signal-level audio tokenizers, such as those employed in this work, often rely on residual quantization to capture information across different frequencies. While effective, this approach faces challenges in managing long token lengths, particularly for high-resolution audio signals. To mitigate this, we adopt a multi-scale approach that reduces token length while maintaining reconstruction quality. However, semantic tokenizers offer a promising alternative by achieving shorter token lengths with higher information density. Integrating semantic information into multi-scale quantized tokens could further reduce token length while enhancing the richness and efficiency of latent representations. Addressing this integration and improving scalability will be a key direction for our future research.

\bibliography{main.bib}

\begin{thebibliography}{45}
\providecommand{\natexlab}[1]{#1}

\bibitem[{Achiam et~al.(2023)Achiam, Adler, Agarwal, Ahmad, Akkaya, Aleman, Almeida, Altenschmidt, Altman, Anadkat et~al.}]{achiam2023gpt}
Josh Achiam, Steven Adler, Sandhini Agarwal, Lama Ahmad, Ilge Akkaya, Florencia~Leoni Aleman, Diogo Almeida, Janko Altenschmidt, Sam Altman, Shyamal Anadkat, et~al. 2023.
\newblock Gpt-4 technical report.
\newblock \emph{arXiv preprint arXiv:2303.08774}.

\bibitem[{Baevski et~al.(2020)Baevski, Zhou, Mohamed, and Auli}]{baevski2020wav2vec}
Alexei Baevski, Yuhao Zhou, Abdelrahman Mohamed, and Michael Auli. 2020.
\newblock wav2vec 2.0: A framework for self-supervised learning of speech representations.
\newblock \emph{Advances in neural information processing systems}, 33:12449--12460.

\bibitem[{Bahdanau et~al.(2014)Bahdanau, Cho, and Bengio}]{bahdanau2014neural}
Dzmitry Bahdanau, Kyunghyun Cho, and Yoshua Bengio. 2014.
\newblock Neural machine translation by jointly learning to align and translate.
\newblock \emph{arXiv preprint arXiv:1409.0473}.

\bibitem[{Caillon and Esling(2021)}]{caillon2021rave}
Antoine Caillon and Philippe Esling. 2021.
\newblock Rave: A variational autoencoder for fast and high-quality neural audio synthesis.
\newblock \emph{arXiv preprint arXiv:2111.05011}.

\bibitem[{Chang et~al.(2022)Chang, Zhang, Jiang, Liu, and Freeman}]{chang2022maskgit}
Huiwen Chang, Han Zhang, Lu~Jiang, Ce~Liu, and William~T Freeman. 2022.
\newblock Maskgit: Masked generative image transformer.
\newblock In \emph{Proceedings of the IEEE/CVF Conference on Computer Vision and Pattern Recognition}, pages 11315--11325.

\bibitem[{Chen et~al.(2020)Chen, Xie, Vedaldi, and Zisserman}]{chen2020vggsound}
Honglie Chen, Weidi Xie, Andrea Vedaldi, and Andrew Zisserman. 2020.
\newblock Vggsound: A large-scale audio-visual dataset.
\newblock In \emph{ICASSP 2020-2020 IEEE International Conference on Acoustics, Speech and Signal Processing (ICASSP)}, pages 721--725. IEEE.

\bibitem[{Chowdhery et~al.(2023)Chowdhery, Narang, Devlin, Bosma, Mishra, Roberts, Barham, Chung, Sutton, Gehrmann et~al.}]{chowdhery2023palm}
Aakanksha Chowdhery, Sharan Narang, Jacob Devlin, Maarten Bosma, Gaurav Mishra, Adam Roberts, Paul Barham, Hyung~Won Chung, Charles Sutton, Sebastian Gehrmann, et~al. 2023.
\newblock Palm: Scaling language modeling with pathways.
\newblock \emph{Journal of Machine Learning Research}, 24(240):1--113.

\bibitem[{D{\'e}fossez et~al.(2022)D{\'e}fossez, Copet, Synnaeve, and Adi}]{defossez2022high}
Alexandre D{\'e}fossez, Jade Copet, Gabriel Synnaeve, and Yossi Adi. 2022.
\newblock High fidelity neural audio compression.
\newblock \emph{arXiv preprint arXiv:2210.13438}.

\bibitem[{Devlin et~al.(2018)Devlin, Chang, Lee, and Toutanova}]{devlin2018bert}
Jacob Devlin, Ming-Wei Chang, Kenton Lee, and Kristina Toutanova. 2018.
\newblock Bert: Pre-training of deep bidirectional transformers for language understanding.
\newblock \emph{arXiv preprint arXiv:1810.04805}.

\bibitem[{Dosovitskiy et~al.(2020)Dosovitskiy, Beyer, Kolesnikov, Weissenborn, Zhai, Unterthiner, Dehghani, Minderer, Heigold, Gelly et~al.}]{dosovitskiy2020image}
Alexey Dosovitskiy, Lucas Beyer, Alexander Kolesnikov, Dirk Weissenborn, Xiaohua Zhai, Thomas Unterthiner, Mostafa Dehghani, Matthias Minderer, Georg Heigold, Sylvain Gelly, et~al. 2020.
\newblock An image is worth 16x16 words: Transformers for image recognition at scale.
\newblock \emph{arXiv preprint arXiv:2010.11929}.

\bibitem[{Esser et~al.(2021)Esser, Rombach, and Ommer}]{esser2021taming}
Patrick Esser, Robin Rombach, and Bjorn Ommer. 2021.
\newblock Taming transformers for high-resolution image synthesis.
\newblock In \emph{Proceedings of the IEEE/CVF conference on computer vision and pattern recognition}, pages 12873--12883.

\bibitem[{Gemmeke et~al.(2017)Gemmeke, Ellis, Freedman, Jansen, Lawrence, Moore, Plakal, and Ritter}]{gemmeke2017audio}
Jort~F Gemmeke, Daniel~PW Ellis, Dylan Freedman, Aren Jansen, Wade Lawrence, R~Channing Moore, Manoj Plakal, and Marvin Ritter. 2017.
\newblock Audio set: An ontology and human-labeled dataset for audio events.
\newblock In \emph{2017 IEEE international conference on acoustics, speech and signal processing (ICASSP)}, pages 776--780. IEEE.

\bibitem[{Hoffmann et~al.(2022)Hoffmann, Borgeaud, Mensch, Buchatskaya, Cai, Rutherford, Casas, Hendricks, Welbl, Clark et~al.}]{hoffmann2022training}
Jordan Hoffmann, Sebastian Borgeaud, Arthur Mensch, Elena Buchatskaya, Trevor Cai, Eliza Rutherford, Diego de~Las Casas, Lisa~Anne Hendricks, Johannes Welbl, Aidan Clark, et~al. 2022.
\newblock Training compute-optimal large language models.
\newblock \emph{arXiv preprint arXiv:2203.15556}.

\bibitem[{Hsu et~al.(2021)Hsu, Bolte, Tsai, Lakhotia, Salakhutdinov, and Mohamed}]{hsu2021hubert}
Wei-Ning Hsu, Benjamin Bolte, Yao-Hung~Hubert Tsai, Kushal Lakhotia, Ruslan Salakhutdinov, and Abdelrahman Mohamed. 2021.
\newblock Hubert: Self-supervised speech representation learning by masked prediction of hidden units.
\newblock \emph{IEEE/ACM transactions on audio, speech, and language processing}, 29:3451--3460.

\bibitem[{Kilgour et~al.(2018)Kilgour, Zuluaga, Roblek, and Sharifi}]{kilgour2018fr}
Kevin Kilgour, Mauricio Zuluaga, Dominik Roblek, and Matthew Sharifi. 2018.
\newblock Fr$\backslash$'echet audio distance: A metric for evaluating music enhancement algorithms.
\newblock \emph{arXiv preprint arXiv:1812.08466}.

\bibitem[{Kong et~al.(2020)Kong, Kim, and Bae}]{kong2020hifi}
Jungil Kong, Jaehyeon Kim, and Jaekyoung Bae. 2020.
\newblock Hifi-gan: Generative adversarial networks for efficient and high fidelity speech synthesis.
\newblock \emph{Advances in neural information processing systems}, 33:17022--17033.

\bibitem[{Kreuk et~al.(2022)Kreuk, Synnaeve, Polyak, Singer, D{\'e}fossez, Copet, Parikh, Taigman, and Adi}]{kreuk2022audiogen}
Felix Kreuk, Gabriel Synnaeve, Adam Polyak, Uriel Singer, Alexandre D{\'e}fossez, Jade Copet, Devi Parikh, Yaniv Taigman, and Yossi Adi. 2022.
\newblock Audiogen: Textually guided audio generation.
\newblock \emph{arXiv preprint arXiv:2209.15352}.

\bibitem[{Kumar et~al.(2019)Kumar, Kumar, De~Boissiere, Gestin, Teoh, Sotelo, De~Brebisson, Bengio, and Courville}]{kumar2019melgan}
Kundan Kumar, Rithesh Kumar, Thibault De~Boissiere, Lucas Gestin, Wei~Zhen Teoh, Jose Sotelo, Alexandre De~Brebisson, Yoshua Bengio, and Aaron~C Courville. 2019.
\newblock Melgan: Generative adversarial networks for conditional waveform synthesis.
\newblock \emph{Advances in neural information processing systems}, 32.

\bibitem[{Kumar et~al.(2024)Kumar, Seetharaman, Luebs, Kumar, and Kumar}]{kumar2024high}
Rithesh Kumar, Prem Seetharaman, Alejandro Luebs, Ishaan Kumar, and Kundan Kumar. 2024.
\newblock High-fidelity audio compression with improved rvqgan.
\newblock \emph{Advances in Neural Information Processing Systems}, 36.

\bibitem[{Lee et~al.(2022)Lee, Kim, Kim, Cho, and Han}]{lee2022autoregressive}
Doyup Lee, Chiheon Kim, Saehoon Kim, Minsu Cho, and Wook-Shin Han. 2022.
\newblock Autoregressive image generation using residual quantization.
\newblock In \emph{Proceedings of the IEEE/CVF Conference on Computer Vision and Pattern Recognition}, pages 11523--11532.

\bibitem[{Li et~al.(2024{\natexlab{a}})Li, Qiu, Chen, Kuen, Gu, Raj, and Lin}]{li2024imagefolder}
Xiang Li, Kai Qiu, Hao Chen, Jason Kuen, Jiuxiang Gu, Bhiksha Raj, and Zhe Lin. 2024{\natexlab{a}}.
\newblock Imagefolder: Autoregressive image generation with folded tokens.
\newblock \emph{arXiv preprint arXiv:2410.01756}.

\bibitem[{Li et~al.(2024{\natexlab{b}})Li, Qiu, Chen, Kuen, Gu, Wang, Lin, and Raj}]{li2024xq}
Xiang Li, Kai Qiu, Hao Chen, Jason Kuen, Jiuxiang Gu, Jindong Wang, Zhe Lin, and Bhiksha Raj. 2024{\natexlab{b}}.
\newblock Xq-gan: An open-source image tokenization framework for autoregressive generation.
\newblock \emph{arXiv preprint arXiv:2412.01762}.

\bibitem[{Li et~al.(2024{\natexlab{c}})Li, Qiu, Chen, Kuen, Lin, Singh, and Raj}]{li2024controlvar}
Xiang Li, Kai Qiu, Hao Chen, Jason Kuen, Zhe Lin, Rita Singh, and Bhiksha Raj. 2024{\natexlab{c}}.
\newblock Controlvar: Exploring controllable visual autoregressive modeling.
\newblock \emph{arXiv preprint arXiv:2406.09750}.

\bibitem[{Liu et~al.(2024{\natexlab{a}})Liu, Yuan, Liu, Mei, Kong, Tian, Wang, Wang, Wang, and Plumbley}]{liu2024audioldm}
Haohe Liu, Yi~Yuan, Xubo Liu, Xinhao Mei, Qiuqiang Kong, Qiao Tian, Yuping Wang, Wenwu Wang, Yuxuan Wang, and Mark~D Plumbley. 2024{\natexlab{a}}.
\newblock Audioldm 2: Learning holistic audio generation with self-supervised pretraining.
\newblock \emph{IEEE/ACM Transactions on Audio, Speech, and Language Processing}.

\bibitem[{Liu et~al.(2024{\natexlab{b}})Liu, Huang, Liu, Cao, Wang, Cheng, Zheng, and Zhao}]{liu2024audiolcm}
Huadai Liu, Rongjie Huang, Yang Liu, Hengyuan Cao, Jialei Wang, Xize Cheng, Siqi Zheng, and Zhou Zhao. 2024{\natexlab{b}}.
\newblock Audiolcm: Text-to-audio generation with latent consistency models.
\newblock \emph{arXiv preprint arXiv:2406.00356}.

\bibitem[{Oord et~al.(2016)Oord, Dieleman, Zen, Simonyan, Vinyals, Graves, Kalchbrenner, Senior, and Kavukcuoglu}]{oord2016wavenet}
Aaron van~den Oord, Sander Dieleman, Heiga Zen, Karen Simonyan, Oriol Vinyals, Alex Graves, Nal Kalchbrenner, Andrew Senior, and Koray Kavukcuoglu. 2016.
\newblock Wavenet: A generative model for raw audio.
\newblock \emph{arXiv preprint arXiv:1609.03499}.

\bibitem[{Radford et~al.(2019)Radford, Wu, Child, Luan, Amodei, Sutskever et~al.}]{radford2019language}
Alec Radford, Jeffrey Wu, Rewon Child, David Luan, Dario Amodei, Ilya Sutskever, et~al. 2019.
\newblock Language models are unsupervised multitask learners.
\newblock \emph{OpenAI blog}, 1(8):9.

\bibitem[{Razavi et~al.(2019)Razavi, Van~den Oord, and Vinyals}]{razavi2019generating}
Ali Razavi, Aaron Van~den Oord, and Oriol Vinyals. 2019.
\newblock Generating diverse high-fidelity images with vq-vae-2.
\newblock \emph{Advances in neural information processing systems}, 32.

\bibitem[{Salimans et~al.(2016)Salimans, Goodfellow, Zaremba, Cheung, Radford, and Chen}]{salimans2016improved}
Tim Salimans, Ian Goodfellow, Wojciech Zaremba, Vicki Cheung, Alec Radford, and Xi~Chen. 2016.
\newblock Improved techniques for training gans.
\newblock \emph{Advances in neural information processing systems}, 29.

\bibitem[{Siddiqui et~al.(2024)Siddiqui, Alliegro, Artemov, Tommasi, Sirigatti, Rosov, Dai, and Nie{\ss}ner}]{siddiqui2024meshgpt}
Yawar Siddiqui, Antonio Alliegro, Alexey Artemov, Tatiana Tommasi, Daniele Sirigatti, Vladislav Rosov, Angela Dai, and Matthias Nie{\ss}ner. 2024.
\newblock Meshgpt: Generating triangle meshes with decoder-only transformers.
\newblock In \emph{Proceedings of the IEEE/CVF Conference on Computer Vision and Pattern Recognition}, pages 19615--19625.

\bibitem[{Siuzdak et~al.(2024)Siuzdak, Gr{\"o}tschla, and Lanzend{\"o}rfer}]{siuzdak2024snac}
Hubert Siuzdak, Florian Gr{\"o}tschla, and Luca~A Lanzend{\"o}rfer. 2024.
\newblock Snac: Multi-scale neural audio codec.
\newblock \emph{arXiv preprint arXiv:2410.14411}.

\bibitem[{Sun et~al.(2024)Sun, Jiang, Chen, Zhang, Peng, Luo, and Yuan}]{sun2024autoregressive}
Peize Sun, Yi~Jiang, Shoufa Chen, Shilong Zhang, Bingyue Peng, Ping Luo, and Zehuan Yuan. 2024.
\newblock Autoregressive model beats diffusion: Llama for scalable image generation.
\newblock \emph{arXiv preprint arXiv:2406.06525}.

\bibitem[{Sutskever et~al.(2014)Sutskever, Vinyals, and Le}]{sutskever2014sequence}
Ilya Sutskever, Oriol Vinyals, and Quoc~V Le. 2014.
\newblock Sequence to sequence learning with neural networks.
\newblock \emph{Advances in neural information processing systems}, 27.

\bibitem[{Tian et~al.(2024)Tian, Jiang, Yuan, Peng, and Wang}]{tian2024visual}
Keyu Tian, Yi~Jiang, Zehuan Yuan, Bingyue Peng, and Liwei Wang. 2024.
\newblock Visual autoregressive modeling: Scalable image generation via next-scale prediction.
\newblock \emph{arXiv preprint arXiv:2404.02905}.

\bibitem[{Touvron et~al.(2023)Touvron, Lavril, Izacard, Martinet, Lachaux, Lacroix, Rozi{\`e}re, Goyal, Hambro, Azhar et~al.}]{touvron2023llama}
Hugo Touvron, Thibaut Lavril, Gautier Izacard, Xavier Martinet, Marie-Anne Lachaux, Timoth{\'e}e Lacroix, Baptiste Rozi{\`e}re, Naman Goyal, Eric Hambro, Faisal Azhar, et~al. 2023.
\newblock Llama: Open and efficient foundation language models.
\newblock \emph{arXiv preprint arXiv:2302.13971}.

\bibitem[{Van Den~Oord et~al.(2017)Van Den~Oord, Vinyals et~al.}]{van2017neural}
Aaron Van Den~Oord, Oriol Vinyals, et~al. 2017.
\newblock Neural discrete representation learning.
\newblock \emph{Advances in neural information processing systems}, 30.

\bibitem[{Vaswani et~al.(2017)Vaswani, Shazeer, Parmar, Uszkoreit, Jones, Gomez, Kaiser, and Polosukhin}]{vaswani2017attention}
Ashish Vaswani, Noam Shazeer, Niki Parmar, Jakob Uszkoreit, Llion Jones, Aidan~N Gomez, {\L}ukasz Kaiser, and Illia Polosukhin. 2017.
\newblock Attention is all you need.
\newblock \emph{Advances in neural information processing systems}, 30.

\bibitem[{Weissenborn et~al.(2019)Weissenborn, T{\"a}ckstr{\"o}m, and Uszkoreit}]{weissenborn2019scaling}
Dirk Weissenborn, Oscar T{\"a}ckstr{\"o}m, and Jakob Uszkoreit. 2019.
\newblock Scaling autoregressive video models.
\newblock \emph{arXiv preprint arXiv:1906.02634}.

\bibitem[{Wu et~al.(2023{\natexlab{a}})Wu, Fan, Liu, Zheng, Gong, Jiao, Li, Guo, Duan, Chen et~al.}]{wu2023ar}
Tong Wu, Zhihao Fan, Xiao Liu, Hai-Tao Zheng, Yeyun Gong, Jian Jiao, Juntao Li, Jian Guo, Nan Duan, Weizhu Chen, et~al. 2023{\natexlab{a}}.
\newblock Ar-diffusion: Auto-regressive diffusion model for text generation.
\newblock \emph{Advances in Neural Information Processing Systems}, 36:39957--39974.

\bibitem[{Wu et~al.(2023{\natexlab{b}})Wu, Chen, Zhang, Hui, Berg-Kirkpatrick, and Dubnov}]{wu2023large}
Yusong Wu, Ke~Chen, Tianyu Zhang, Yuchen Hui, Taylor Berg-Kirkpatrick, and Shlomo Dubnov. 2023{\natexlab{b}}.
\newblock Large-scale contrastive language-audio pretraining with feature fusion and keyword-to-caption augmentation.
\newblock In \emph{ICASSP 2023-2023 IEEE International Conference on Acoustics, Speech and Signal Processing (ICASSP)}, pages 1--5. IEEE.

\bibitem[{Yang et~al.(2023{\natexlab{a}})Yang, Liu, Huang, Tian, Weng, and Zou}]{yang2023hifi}
Dongchao Yang, Songxiang Liu, Rongjie Huang, Jinchuan Tian, Chao Weng, and Yuexian Zou. 2023{\natexlab{a}}.
\newblock Hifi-codec: Group-residual vector quantization for high fidelity audio codec.
\newblock \emph{arXiv preprint arXiv:2305.02765}.

\bibitem[{Yang et~al.(2023{\natexlab{b}})Yang, Yu, Wang, Wang, Weng, Zou, and Yu}]{yang2023diffsound}
Dongchao Yang, Jianwei Yu, Helin Wang, Wen Wang, Chao Weng, Yuexian Zou, and Dong Yu. 2023{\natexlab{b}}.
\newblock Diffsound: Discrete diffusion model for text-to-sound generation.
\newblock \emph{IEEE/ACM Transactions on Audio, Speech, and Language Processing}, 31:1720--1733.

\bibitem[{Yu et~al.(2021)Yu, Li, Koh, Zhang, Pang, Qin, Ku, Xu, Baldridge, and Wu}]{yu2021vector}
Jiahui Yu, Xin Li, Jing~Yu Koh, Han Zhang, Ruoming Pang, James Qin, Alexander Ku, Yuanzhong Xu, Jason Baldridge, and Yonghui Wu. 2021.
\newblock Vector-quantized image modeling with improved vqgan.
\newblock \emph{arXiv preprint arXiv:2110.04627}.

\bibitem[{Zeghidour et~al.(2021)Zeghidour, Luebs, Omran, Skoglund, and Tagliasacchi}]{zeghidour2021soundstream}
Neil Zeghidour, Alejandro Luebs, Ahmed Omran, Jan Skoglund, and Marco Tagliasacchi. 2021.
\newblock Soundstream: An end-to-end neural audio codec.
\newblock \emph{IEEE/ACM Transactions on Audio, Speech, and Language Processing}, 30:495--507.

\bibitem[{Zhang et~al.(2018)Zhang, Goodfellow, Metaxas, and Odena}]{zhang2018self}
Han Zhang, Ian Goodfellow, Dimitris Metaxas, and Augustus Odena. 2018.
\newblock Self-attention generative adversarial networks.
\newblock In \emph{International Conference on Machine Learning}, pages 7354--7363.

\end{thebibliography}

\clearpage


\appendix
\begin{algorithm}[]
\caption{Multi-scale residual quantization}
\label{alg:multi-scale-encodec}
\textbf{Input}: Raw Audio Signal $\mathcal{A}$\\
\textbf{Parameter}: Encoder $\mathcal{E}$, Decoder $\mathcal{D}$, Quantizer $\mathcal{Q}_{i=1}^{K}$\\
\textbf{Hyper-parameter}: Resolution $(t_k)_{k=1}^{K}$, interpolation $\phi$
\begin{algorithmic}[1] 
\STATE $f=\mathcal{E}(\mathcal{A})$
\STATE $R=[]$
\FOR{$k$ in $(1,2,...K)$}
\IF {$k = K$}
\STATE $r_k = \mathcal{Q}_k(f)$
\STATE $z_k = \textit{lookup}(Q_k, r_k)$
\ELSE
\STATE $r_k = \mathcal{Q}_k(\textit{interpolate}(f, t_k))$
\STATE $z_k = \textit{lookup}(Q_k, r_k)$
\STATE $z_k = \textit{interpolate}(z_k, t_K)$
\ENDIF
\STATE $R = R + r_k$
\STATE $f = f - \phi(z_k)$
\ENDFOR
\STATE \textbf{return} R
\end{algorithmic}
\end{algorithm}

\section{Supplementary material}


\begin{algorithm}[t]
\caption{Multi-scale AR Generation}
\label{alg:multi-scale-AR}
\textbf{Input}: Text $\mathcal{T}$\\
\textbf{Parameter}: Decoder $\mathcal{D}$, GPT $AR$, conditional Model $C$\\
\textbf{Hyper-parameter}: Resolution $(t_k)_{k=1}^{K}$, interpolation $\phi$
\begin{algorithmic}[1] 
\STATE $x_0= C(T)$
\STATE $R=[], S=[x_0]$
\FOR{$k$ in $(1,2,...K)$}
\STATE $x_k = AR(S)$
\STATE $R = R + x_k$
\IF {$k = K$}
\STATE break
\ELSE
\STATE $x_k = interpolate(x_k, t_K)$
\STATE $x_k = \phi (interpolate(x_k, t_{k+1}))$
\ENDIF
\STATE $S = S + x_k$
\ENDFOR
\STATE $A = D(R)$
\STATE \textbf{return} A
\end{algorithmic}
\end{algorithm}

\subsection{Multi-Scale Residual Quantization}
Multi-scale residual quantization (MSRQ) in our SAT is designed to efficiently encode audio signals by leveraging multiple quantization stages on different scales. Specifically, the MSRQ process begins with the raw audio signal being encoded into a feature representation. This representation is then passed through a series of quantizers, each corresponding to a different scale. For each scale, the feature map is downsampled to match the target resolution before quantization. After quantization, the feature map is upsampled back to its original resolution. Due to information loss in interpolation, the upsampled feature map is further processed through our upsampling network to recover the information for each scale. The residual error, calculated after each quantization step, is passed to the next quantizer, allowing the model to refine the audio representation iteratively. The pesudo-code implementation can be shown in \cref{alg:multi-scale-encodec}.

\subsection{Scale-level Acoustic AutoRegressive Generation.}
Our AAR method begins by taking the scale-level tokens generated by the MSRQ process. These tokens are organized hierarchically, with each scale capturing different levels of detail in the audio signal—from coarse, low-frequency information to fine, high-frequency details. In generation, the process is structured to predict these scales sequentially, starting from the coarsest scale and progressing to finer scales. As \cref{alg:multi-scale-AR} illustrated, our AAR first initializes the generation process by producing an initial latent representation from the input text using a conditional model. This initial representation serves as the starting point for the autoregressive prediction. The model then proceeds through each scale, beginning with the coarsest, and generates the corresponding tokens by conditioning on the sequence of tokens generated thus far. After each scale's tokens are predicted, they are interpolated and refined to align with the resolution of the next finer scale. This iterative process continues until all scales are generated, ensuring a smooth and coherent progression from low to high-frequency details. Finally, the aggregated tokens from all scales are decoded into a complete audio signal, resulting in a high-fidelity output that effectively captures the nuances of the original audio.

\subsection{Scale Scheduling}
\begin{figure}[!h]
  \centering
    \includegraphics[width=\linewidth]{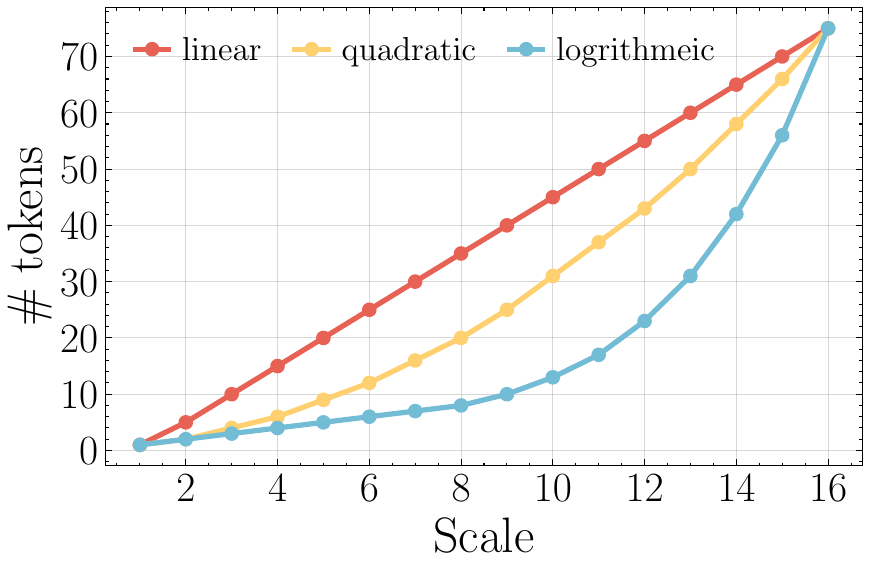}
    \vspace{-0.3cm}
  \caption{Visualization of Linear, Quadratic, and Logarithmic scale scheduling across the range from 1 to 75.}
  \vspace{-0.2cm}
  \label{fig:scale}
\end{figure}

In our paper, we explore three types of scale scheduling: Linear, Quadratic, and Logarithmic. Specifically, Linear scheduling ensures that the difference between each scale is consistent and linear. For example, in our linear scheduling approach, we start from a scale of 1 and increase to 75 using 16 scales, resulting in a difference of approximately 5 between each consecutive scale. The detail visualization of our scale setting can be shown in \cref{fig:scale}.

\subsection{Codebook Utilization}
We analyze the codebook utilization across different models. In particular, we observe that the codebook utilization for Encodec \cite{defossez2022high} consistently reaches $99\%$, indicating that the entire codebook is actively used during the encoding process. In contrast, our SAT model exhibits a lower utilization rate. Specifically, we find that the codebook utilization in every scale of the SAT model remains at approximately $60\%$. We hypothesize that this discrepancy is caused by the inherent structure of the SAT model, where each time we downsample the input and apply quantization, the model becomes increasingly selective in its use of the codebook entries. 


\subsection{Training Cost Analysis}
Our AAR exhibits strong performance in terms of latency and quality while also efficiently reducing the training cost of the model. To be more specific, the vanilla AR achieved an FAD of 8.07 with a classifier-free guidance scale of 4 after training for 100 epochs, while our AAR achieved an improved FAD of 5.70 under the same settings in just 45 epochs. As shown in \cref{tab:train_cost}, to achieve the same capacity of vanilla AR, our AAR only needs to train 20 epochs and efficiently save approximately $80\%$ training cost.

\tabtrain{}

\begin{figure}[]
  \centering
  \includegraphics[width=\linewidth]{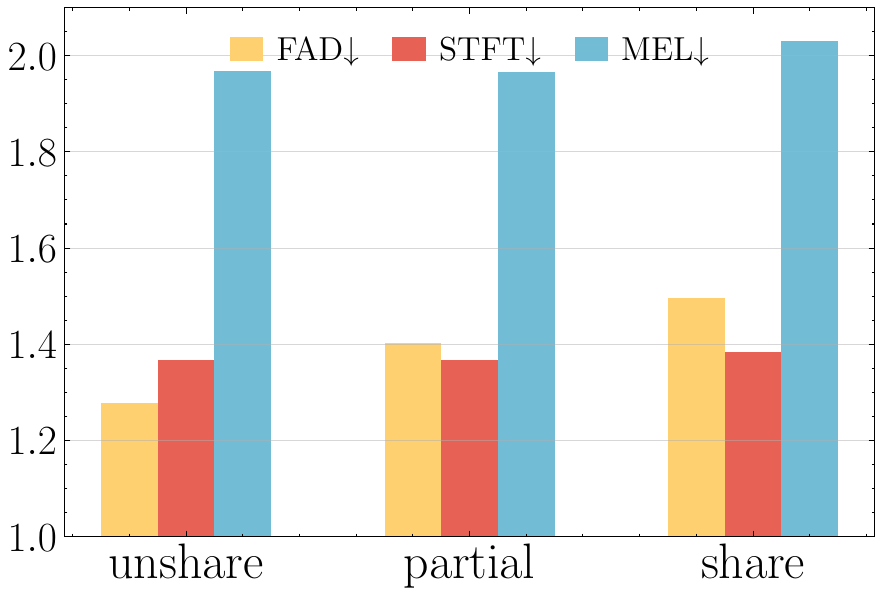}
  \caption{Ablation study of upsampling functions on SAT.}
  \label{fig:exp4}
\end{figure}

\begin{figure}[]
  \centering
    \vspace{-0.2cm}
    \includegraphics[width=\linewidth]{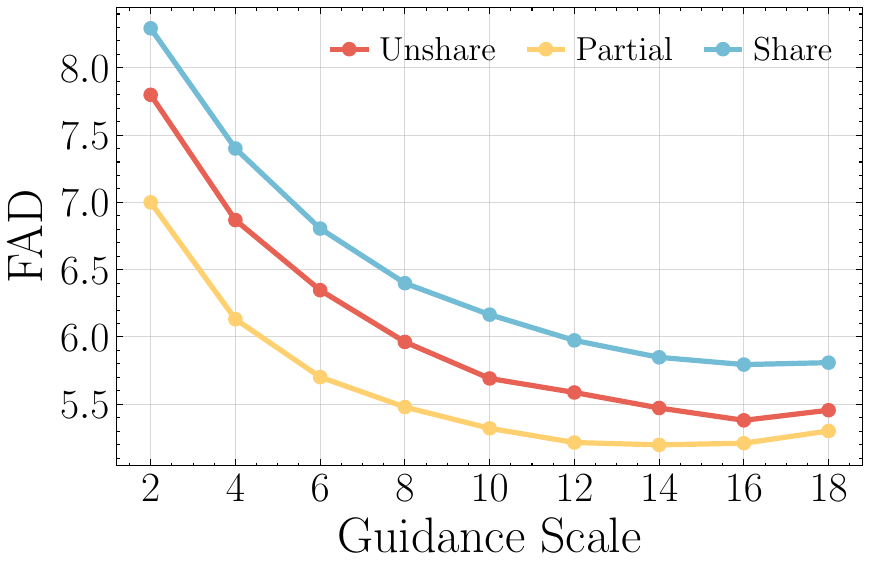}
    \vspace{-0.7cm}
  \caption{Ablation study of upsampling functions on AAR.}
  \vspace{-0.2cm}
  \label{fig:exp5}
\end{figure}
\subsection{Upsampling Function}
In our work, to efficiently recover information loss from downsampling, we use a 1D convolutional layer after vanilla upsampling to ensure unique information on each scale is preserved and accurately represented. We evaluated its effectiveness through three configurations: unshared (each quantizer has its own convolutional layer); partially shared (approximately three quantizers share one layer); and fully shared (all quantizers use the same layer) to validate the effectiveness of this approach in distinguishing and splitting information across different scales for multi-scale reconstruction, and so on, generation. Our experiments (see \cref{fig:exp4}) show that the performance of unshared, partially shared, and fully shared networks in reconstruction is similar, indicating that all configurations effectively maintain audio quality during reconstruction. However, their impact on generation can be seen in \cref{fig:exp5}, where the partially shared architecture significantly improves generation quality. 

\clearpage

\end{document}